\documentstyle[12pt,epsfig]{article}
\catcode`\@=11
\renewcommand{\thefootnote}{\fnsymbol{footnote}}

\newcounter{sectionc}\newcounter{subsectionc}\newcounter{subsubsectionc}
\renewcommand{\section}[1] {\vspace*{0.6cm}\addtocounter{sectionc}{1}
\setcounter{subsectionc}{0}\setcounter{subsubsectionc}{0}\noindent
	{\normalsize\bf\thesectionc. #1}\par\vspace*{0.4cm}}
\renewcommand{\subsection}[1] {\vspace*{0.6cm}\addtocounter{subsectionc}{1}
	\setcounter{subsubsectionc}{0}\noindent
	{\normalsize\it\thesectionc.\thesubsectionc. #1}\par\vspace*{0.4cm}}
\renewcommand{\subsubsection}[1]
{\vspace*{0.6cm}\addtocounter{subsubsectionc}{1}
	\noindent {\normalsize\rm\thesectionc.\thesubsectionc.\thesubsubsectionc.
	#1}\par\vspace*{0.4cm}}

\newcounter{appendixc}
\newcounter{subappendixc}[appendixc]
\newcounter{subsubappendixc}[subappendixc]

\renewcommand{\appendix}[1] {\vspace*{0.6cm}
        \refstepcounter{appendixc}
        \setcounter{figure}{0}
        \setcounter{table}{0}
        \setcounter{equation}{0}
        \renewcommand{\thefigure}{\Alph{appendixc}.\arabic{figure}}
        \renewcommand{\thetable}{\Alph{appendixc}.\arabic{table}}
        \renewcommand{\theappendixc}{\Alph{appendixc}}
        \renewcommand{\theequation}{\Alph{appendixc}.\arabic{equation}}
        \noindent{\bf Appendix \theappendixc #1}\par\vspace*{0.4cm}}

\def\abstracts#1{{
	\centering{\begin{minipage}{12.2truecm}\footnotesize%
\baselineskip=12pt\noindent
	\centerline{\footnotesize ABSTRACT}\vspace*{0.3cm}
	\parindent=0pt #1
	\end{minipage}}\par}}

\renewenvironment{thebibliography}[1]
	{\begin{list}{\arabic{enumi}.}
	{\usecounter{enumi}\setlength{\parsep}{0pt}
\setlength{\leftmargin 1.25cm}{\rightmargin 0pt}
	 \setlength{\itemsep}{0pt} \settowidth
	{\labelwidth}{#1.}\sloppy}}{\end{list}}

\newcounter{itemlistc}
\newcounter{romanlistc}
\newcounter{alphlistc}
\newcounter{arabiclistc}

\newcommand{\fcaption}[1]{
        \refstepcounter{figure}
        \setbox\@tempboxa = \hbox{\footnotesize Fig.~\thefigure. #1}
        \ifdim \wd\@tempboxa > 6in
           {\begin{center}
        \parbox{6in}{\footnotesize\baselineskip=12pt Fig.~\thefigure. #1}
            \end{center}}
        \else
             {\begin{center}
             {\footnotesize Fig.~\thefigure. #1}
              \end{center}}
        \fi}

\newcommand{\tcaption}[1]{
        \refstepcounter{table}
        \setbox\@tempboxa = \hbox{\footnotesize Table~\thetable. #1}
        \ifdim \wd\@tempboxa > 6in
           {\begin{center}
        \parbox{6in}{\footnotesize\baselineskip=12pt Table~\thetable. #1}
            \end{center}}
        \else
             {\begin{center}
             {\footnotesize Table~\thetable. #1}
              \end{center}}
        \fi}

\def\@citex[#1]#2{\if@filesw\immediate\write\@auxout
	{\string\citation{#2}}\fi
\def\@citea{}\@cite{\@for\@citeb:=#2\do
	{\@citea\def\@citea{,}\@ifundefined
	{b@\@citeb}{{\bf ?}\@warning
	{Citation `\@citeb' on page \thepage \space undefined}}
	{\csname b@\@citeb\endcsname}}}{#1}}

\newif\if@cghi
\def\cite{\@cghitrue\@ifnextchar [{\@tempswatrue
	\@citex}{\@tempswafalse\@citex[]}}
\def\citelow{\@cghifalse\@ifnextchar [{\@tempswatrue
	\@citex}{\@tempswafalse\@citex[]}}
\def\@cite#1#2{{$\null^{#1}$\if@tempswa\typeout
	{IJCGA warning: optional citation argument
	ignored: `#2'} \fi}}

 1
 1
 1

\font\ninerm=cmr9

\catcode`\@=11
\long\def\@makefntext#1{
\protect\noindent \hbox to 3.2pt {\hskip-.9pt
$^{{\ninerm\@thefnmark}}$\hfil}#1\hfill}		%CAN BE USED

\def\@makefnmark{\hbox to 0pt{$^{\@thefnmark}$\hss}}  %ORIGINAL

\def\ps@myheadings{\let\@mkboth\@gobbletwo
\def\@oddhead{\hbox{}
\rightmark\hfil\ninerm\thepage}
\def\@oddfoot{}\def\@evenhead{\ninerm\thepage\hfil
\leftmark\hbox{}}\def\@evenfoot{}
\def\sectionmark##1{}\def\subsectionmark##1{}}

\renewcommand{\[}{\begin{equation}}
\renewcommand{\]}{\end{equation}}

\newcommand{\fps}{{\mathrm{fps}}}
\newcommand{\nbar}{\bar n}        % NB parameters: must be used in math mode
\newcommand{\Nbar}{\bar N}        % only, or it won't work!
\newcommand{\nc}{\bar n_c}

\newcommand{\jetset}{{\sc Jetset}}

\newcommand{\NF}{{\cal{N}}_{\kern -1.9pt f}}     %math mode only
\newcommand{\NC}{{\cal{N}}_{\kern -1.7pt c}}     %math mode only
\newcommand{\pT}{{p\kern -.2pt\lower 4pt\hbox{\tiny T}}}    %works?
\newcommand{\pL}{{p\kern -.2pt\lower 4pt\hbox{\tiny L}}}    %works?
\newcommand{\Dy}{\Delta y}

\newcommand{\eref}[1]{(\ref{#1})}      % automatically puts ( ) around ref.s

\newif\ifpreprint
\preprinttrue

\begin{document}
\ifpreprint{\begin{flushright}
  LU TP 95-28\\DFTT 65/95\\MPI-PhT/95-106\\
  November 2nd, 1995\end{flushright}}\vfill\fi
\centerline{\normalsize\bf CLAN STRUCTURE IN RAPIDITY INTERVALS
  \ifpreprint\else\footnote{talk presented by R. Ugoccioni}\fi}
\baselineskip=16pt

%\vfill
\ifpreprint\vspace{2.5cm}\else\vspace*{0.2cm}\fi
\centerline{\footnotesize ROBERTO UGOCCIONI}
\baselineskip=13pt
\centerline{\footnotesize\it Department of Theoretical Physics,
Lund University}
\baselineskip=12pt
\centerline{\footnotesize\it S\"olvegatan 14 A, S-22362 Lund, Sweden}
%\centerline{\footnotesize E-mail: roberto@thep.lu.se}
\vspace*{0.3cm}
%\centerline{\footnotesize and}
%\vspace*{0.3cm}
\centerline{\footnotesize ALBERTO GIOVANNINI}
\baselineskip=13pt
\centerline{\footnotesize\it Dipartimento di Fisica Teorica,
Universit\`a di Torino and I.N.F.N. -- Sezione di Torino}
\baselineskip=12pt
\centerline{\footnotesize\it via P. Giuria 1, I-10125 Torino, Italy}
\vspace*{0.3cm}
\centerline{\footnotesize and}
\vspace*{0.3cm}
\centerline{\footnotesize SERGIO LUPIA}
\baselineskip=13pt
\centerline{\footnotesize\it Max-Planck-Institut f\"ur Physik,
Werner-Heisenberg-Institut}
\baselineskip=12pt
\centerline{\footnotesize\it F\"ohringer Ring 6, D-80805 M\"unchen, Germany}

%\vfill
\ifpreprint\vspace{2.5cm}\else\vspace*{0.9cm}\fi
\abstracts{\ifpreprint\normalsize\fi
We present a cascading model for a single jet,
inspired to QCD and
to the phenomenological analysis of multiplicity distributions. The model,
describing as it does a two dimensional evolution in virtuality and
rapidity, allows analytical predictions for clan analysis parameters
to be made.}
\ifpreprint\vfill \begin{center}\footnotesize
Talk presented by R. Ugoccioni\\
at the XXV International Symposium\\
on Multiparticle Dynamics,\\
Star\'a Lesn\'a, Slovakia, 12-16 Sept. 1995.\\
To be published in the Proceedings.
\end{center}\newpage\setcounter{page}{1}\pagestyle{plain}\fi
%\vspace*{0.6cm}
\normalsize\baselineskip=15pt
\setcounter{footnote}{0}
\renewcommand{\thefootnote}{\alph{footnote}}

%--------------------------------------------
\section{Introduction}\label{Intro}

Since its proposal\cite{AGLVH:1},
clan analysis has been widely used experimentally in order to interpret
the negative binomial (NB) regularity and as a tool to classify different
reactions\cite{Buschbeck}. It is especially important to stress the
presence of the NB regularity when the multiplicity distributions (MD)
are studied in intervals of rapidity, where conservation laws don't
play a dominant role and therefore an understanding of the dynamics
is most interesting and desirable.

Recently clan analysis has been extended in three directions: the regularity
is better satisfied in the case of single jets, as is suggested by Monte
Carlo studies\cite{Single} and experimentally by the observation
that separating the 2-, 3- and 4-jet sample of events restores the regularity
that was violated (in the shape of the multiplicity distribution,
but not in the clan structure parameters) in the
full sample of events\cite{DEL:4}.
Secondly, one should remember that, in going from the hadronic level
to the partonic one via generalized local parton-hadron duality\cite{AGLVH:2},
the NB regularity is better satisfied. This suggests
that an explanation of the regularity should be sought at partonic level.
The third extension comes by recognizing
that the clan interpretation of the negative binomial distribution
(NBD) is actually more general than the
NBD itself, and leads to the idea that the emission of partons
in a single jet is a two-step process\cite{GSPS}.

The difficulty of analytical calculations in  QCD, and a guess on the
real   nature of clans in multiparticle production
(they might very well be true physical objects)
has prompted us to develop an analytical parton shower
model for a single jet inspired to QCD and at the
same time close to the phenomenological observations.

In this paper we first discuss some properties of two-step processes
in general,
and once this framework has been established, we describe the
model and its results.

%--------------------------------------------
\section{Two-step processes}\label{duepassi}

Assume that the parton
production process involves two independent steps: in the first
step $N$ objects ($N=0,1\dots$), (which we call {\em clans},
and the clans known from the NBD are now a particular case),
are produced with probability $p_N$ and generating function $f(z)$
(in all quantities in this Section
a dependence on the jet energy is understood).
In the second step, each clan produces partons with probability
$q_{n_i}$ (with $i$ labeling the clans: $i=1,2\dots N$) and
generating function $g(z)$.
Since clans are identified by the final partons they generate,
requiring that each clan produce at least one parton (i.e, $q_0=0$)
makes the number of clans unambiguously defined.
Notice that we do not assume that all clans are created equal:
the multiplicity distribution (MD) of an individual clan
can depend on a set of parameters $\xi$ as $\tilde q_{n_i}(\xi)$:
if the value of $\xi$ for a
clan is independent of that of other clans and on the number of clans,
one can define an {\it average clan}:
\[
  q_{n_i} = \int \tilde q_{n_i}(\xi) \phi(\xi) d\xi\>, \label{eq:avgclan}
\]
where $\phi(\xi)$ is the p.d.f.\ for producing a clan with that values
of the parameters. Indeed this property will be used in Sec.~4.

With these assumptions, the final partons MD, $P_n$, has generating function
\[
  F(z) \equiv \sum_{n=0}^\infty z^n P_n = f\left( \>g(z)\> \right)\>.
\]
If we turn now to look at the production in intervals of phase space,
it is clear that the only dependence on the particular interval belongs
to the second step, which is the step that produces the final partons.
Choosing for definiteness to look at a rapidity interval which will be
denoted $\Dy$, we find that each clan produces partons according to
\[
  g(z,\Dy) = \sum_{n=0}^\infty z^n q_n(\Dy) \>,\qquad\qquad q_0(\Dy) \not= 0
  \label{eq:step2}
\]
where the fact that it is possible that a clan produces zero partons
inside $\Dy$ has been emphasized (see also Fig.~\ref{fig:scheme}).
Obviously the final distribution is then given by
\[
  F(z,\Dy) = f\left( \>g(z,\Dy)\> \right).\label{eq:F}
\]
The parameters of interest in this paper are then the average number of
clans, $\Nbar$, and the average number of partons per clan, $\nc^{(0)}$:
\[
  \Nbar \equiv \left.\frac{d f(z)}{dz}\right|_{z=1} \qquad\qquad
  \nc^{(0)}(\Dy) \equiv \left.\frac{d g(z,\Dy)}{dz}\right|_{z=1} =
                  \frac{\nbar(\Dy)}{\Nbar}
\]
where $\nbar(\Dy)$ is the average number of final partons from
Eq.~(\ref{eq:F}).

\subsection{Binomial convolution}

$f(z)$ and $g(z,\Dy)$ can be redefined
so that only clans which produce at least one parton in
the interval $\Dy$ are considered: one simply
subtracts $q_0(\Dy)$ and rescales the distribution correspondingly:
\[
  g_1(z,\Dy) = \frac{g(z,\Dy) - q_0(\Dy)}{1 - q_0(\Dy)}\>.
\]
This implies then
\[
  f_1(z) = f\Bigl( [1-q_0(\Dy)] z + q_0(\Dy) \Bigr)\>.
\]
It can be shown that this is equivalent to the condition that the
probability  that $m$ of the $N$ produced clans generate at least
one parton in $\Dy$ is given by a binomial distribution of
parameter $[1-q_0(\Dy)]$. In particular, the
average number of clans contributing to the interval $\Dy$ is in general
\[
  \Nbar(\Dy) = [1-q_0(\Dy)] \Nbar   \label{eq:NDy}
\]
and therefore one obtains
\[
  \nc^{(1)}(\Dy) = \frac{\nbar(\Dy)}{\Nbar(\Dy)} =
                  \frac{\nc^{(0)}(\Dy)}{[1-q_0(\Dy)]} \>.
\]
\smallskip
\noindent {\it Example 1A.} Suppose that $f(z)$ is a Poisson distribution:
\[
  f(z) = e^{\Nbar(z-1)};
\]
then the redefined distribution is still a Poisson distribution
with parameter given by Eq.~\eref{eq:NDy}:
\[
  F(z,\Dy) = \exp \{ \Nbar(\Dy) [g_1(z,\Dy)-1] \}\>.
\]

\noindent {\it Example 1B.} Consider the case in which $f(z)$ is a shifted
Poisson distribution:
\[
  f(z) = z e^{\Nbar(z-1)};
\]
(such a case will be encountered in Sec.~3). Then the result is
more complex:
\[
  F(z,\Dy) = \Bigl\{ \bigl[ 1-q_0(\Dy) \bigr] g_1(z,\Dy) + q_0(\Dy) \Bigr\}
              e^{ \Nbar(\Dy) [ g_1(z,\Dy)-1 ] }\>.
\]

\subsection{Compound Poisson distribution}

Alternatively, one can try and redefine the generating functions so that the
final MD is a compound Poisson distribution; this is an implicit
assumption made when one makes a fit with a NBD. We want to solve:
\begin{eqnarray}
    F(z,\Dy) &=& \exp \left\{\lambda(\Dy) [g_2(z,\Dy) - 1] \right\} \\
    g_2(0,\Dy) &=& 0
\end{eqnarray}
for $\lambda(\Dy)$ and $g_2(z,\Dy)$ and we obtain:
\begin{eqnarray}
  \lambda(\Dy) &=& - \log\bigl[ F(0,\Dy) \bigr] \\
  g_2(z,\Dy) &=& 1 + { \log\bigl[ F(z,\Dy) \bigr]\over\lambda(\Dy) }\>.
\end{eqnarray}
Here $g_2(z,\Dy)$ is a true probability generating
function $(d^ng_2/dz^n \ge 0 ~\forall n)$ if and only if all
the combinants of the distribution $F$ are positive.
In other words, in some cases it may not be possible to carry out this
redefinition. In case it is, the parameters of interest are $\lambda(\Dy)$
as derived above and
\[
  \nc(\Dy) = \frac{\nbar(\Dy)}{\lambda(\Dy)} =
             \frac{\Nbar\>\nc^{(0)}(\Dy) }{\lambda(\Dy)}\>.
\]

\smallskip
\noindent {\it Example 2A.} For a Poissonian $f(z)$ we find the same
result as in example {\it 1A}:
\begin{eqnarray}
  \lambda(\Dy) &=& \bigl[ 1-q_0(\Dy) \bigr] \Nbar   \\
  g_2(z,\Dy) &=& \frac{g(z,\Dy) - q_0(\Dy) }{ 1 - q_0(\Dy)}
\end{eqnarray}
so that the two transformations coincide. In particular, $\lambda(\Dy)$
is equal to the average number of clans in the interval $\Dy$.

\noindent {\it Example 2B.} For a shifted Poisson, on the
other hand, they do not coincide:
\begin{eqnarray}
 \lambda(\Dy) &=& \bigl[ 1-q_0(\Dy) \bigr] \Nbar - \log[ q_0(\Dy) ]  \\
 g_2(z,\Dy) &=& 1 + \frac{ \Nbar \bigl[ g(z,\Dy)-1 \bigr]  +
        \log\bigl[ g(z,\Dy) \bigr] }{ \lambda(\Dy) }
\end{eqnarray}
and $\lambda(\Dy)$ is not equal to $\Nbar(\Dy)$. In particular it should be
noted that in full phase space (fps), $\Nbar(\fps) = \Nbar$ but
$\lambda(\fps)$ goes to infinity.

\pagebreak[3]

%--------------------------------------------
\section{The GSPS model}\label{model}

\begin{figure}
\begin{center}\hspace{-1cm}\mbox{\epsfig{file=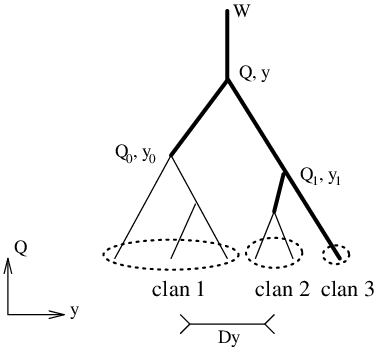}}\end{center}
\fcaption{Schematic representation of the GSPS model. Thick lines
indicate clans creation (step 1) and thin lines indicate
cascading into final partons (step 2). Notice that only clan 1 and 2
produce partons in the interval Dy shown.}\label{fig:scheme}
\end{figure}

The GSPS model is introduced in order to solve analytically
a two dimensional parton evolution in a single jet\cite{GSPS,GSPS:2}.
It describes an ancestor parton,
which degrades in virtuality, and which we follow in rapidity,
 emitting clans of different virtuality and rapidity;
each clan emits partons in a two dimensional cascade process,
see Fig.~\ref{fig:scheme}.

\subsection{Step One}

In Fig.~\ref{fig:scheme} the first step is shown with thick lines.
Because clans are by assumption emitted independently in a cascade fashion,
we take the two branches in each splitting of the ancestor to be
independently regulated by the same splitting function, for which
a form inspired by QCD,
suitably normalized by a Sudakov form factor term, has been chosen:
\[
p_A(Q_0|Q) dQ_0 = p_0^A(Q) C_A(Q) dQ_0 = C_A(Q) \frac{d}{dQ_0}
\left( \frac{1}{C_A(Q_0)} \right) dQ_0 =
d\left(\frac{\log Q_0}{\log Q}\right)^A \;.
\label{psplit}
\]
The parameter $A>0$ thus controls the branching of the ancestor: to
a larger value of $A$ corresponds more branching.

In order to apply this factorization, we have to allow for
local
violations of the energy-momentum conservation law, still requiring
its global validity, i.e.,
offspring partons of virtualities $Q_i$ can fluctuate according to:
\[
Q_0 + Q_1 \not\le Q \ \ , \quad 1 \ \hbox{\rm GeV}\ \le
Q_i \le Q \quad \hbox{\rm [i=0,1]} \; .
\label{weaken}
\]
Of course, constraints on the energy fraction carried away by daughter partons
are also no longer valid, i.e., $z_0+z_1 \not=1$, and
kinematic bounds in rapidity are loosened:
\[
|y_i - y| \le \log {Q \over Q_i} \qquad \hbox{[i=0,1]} \; .
\label{42}
\]
This new condition modifies also the splitting kernel in $z$ which now is
decoupled:
\[
  P(z_0,z_1) dz_0 dz_1 \propto {dz_0 \over z_0}{dz_1 \over z_1} \; .
\label{43}
\]
Notice that, for each branch,
we have taken the singular part of the Altarelli-Parisi
kernel.

Finally, let us mention that the rapidity of
the first splitting is obtained from the energy $W$ and the virtuality
$Q$ by the simple kinematic relationship:
\[
  y = \tanh^{-1}\sqrt{1-Q^2/W^2}.
\]

\subsection{Step Two}

At this point, clans have been emitted with definite initial virtualities and
rapidities. The simplest and most economic assumption is now that inside
each clan a cascade process develops in a way very similar to what has
been described for the first step: each parton branches into two
independent partons. The same form of splitting functions will be applied
as in Sec.~3.1, but with a different parameter:
\[
  p_a(Q_0|Q) dQ_0 = d\left(\frac{\log Q_0}{\log Q}\right)^a \; ;
\]
to a larger value of $a$ corresponds then a larger number of branchings,
and therefore of final partons.
The rapidity part is treated according to Eq.~\eref{42} and \eref{43}.

Finally, one should note that, if energy conservation were strictly
enforced, a parton with virtuality less than 2 GeV could not split.
Because of Eq.~(\ref{weaken}) this is no longer necessary, but we
keep it nonetheless as our cut-off procedure.

%--------------------------------------------
\section{The structure of the calculation}\label{calcstru}

The calculation has been described in detail in~\cite{GSPS:2},
here we will only outline its structure.

Since, according to the model, clans of given virtuality and rapidity
are produced, one must first calculate the generating function for the
MD of a clan with initial virtuality $Q$ and rapidity $y$ to produce
partons in the interval $\Dy$, which we call $g(z,\Dy,Q,y)$.
It would correspond to Eq.~\eref{eq:step2} in Sect.~2.
It can be done based on step two of the model only.

We then calculate the probability to emit a clan with initial virtuality
$Q$ and initial rapidity $y$, which we call $\phi(Q,y)$.
This can be done based on step one of the model
only.

Finally one defines the generating function for the MD of an average
clan similarly to what is done in Eq.~\eref{eq:avgclan}:
\[
  g(z,\Dy) = \int_1^W dQ \int dy\> g(z,\Dy,Q,y) \phi(Q,y) .
\]
This distribution is then to be convoluted with the generating
function for the number of clans, $f(z,W)$, which depends of course
only on step one of the model.
\medskip

\noindent {\it Example.} It is perhaps best to illustrate
this procedure by an example, which
for simplicity only regards full phase space\cite{AGLVH:4}.
Let us examine a pure birth model for a single clan: the MD is
a shifted geometrical distribution
\[
  g(z,\fps,Q,y) = \frac{z}{z - \nu(Q) (z-1)}
\]
where the average value $\nu(Q)$ is assumed (for the sake of
this example only: in the GSPS model its explicit form can
be calculated analytically) to be a specific
function of the clan virtuality $Q$:
\[
  \phi(Q) dQ \propto \frac{d\nu}{\nu}.
\]
The MD of an average clan is given by
\[
  g(z) = \frac{1}{\log\nu} \int_1^\nu\frac{z}{z - \nu' (z-1)}
           \frac{d\nu'}{\nu'} = \frac{\log(1-bz)}{\log(1-b)}
\]
where $b=1-1/\nu$. It is a logarithmic distribution which,
when  convoluted with a Poisson distribution for the first step, gives a NBD.
It should be pointed out that this example does not apply directly
to the GSPS: as shown in the next section, the MD of an average clan is not
a logarithmic distribution (but resembles it); the MD for clans is not
a Poisson distribution, but a shifted Poisson distribution.

%--------------------------------------------
\section{Results and discussion}\label{discuss}

The model can be solved analytically under a few mild mathematical
approximations: the full solution in rapidity intervals is however
far from being simple and compact. The results are therefore
presented in graphical form, but in order to get the general flavour
it is interesting to quote the analytical result in full phase space:
the MD of final clans is given by
\[
  f(z,W) = z \exp\Bigl\{[\Nbar(W) - 1][z - 1]\Bigr\}\>;
  \label{eq:ffps}
\]
because at least one parton (the ancestor) is always present in the cascade,
the total distribution of clans turns out to be a shifted Poisson
distribution. The distribution inside an average clan is given by
\[
  g(z,W) = \frac{1}{\Nbar(W)}\left\{ z+\frac{A}{a}u_a(W) -
    \frac{A}{a}\log\left[z+(1-z)\exp\{u_a(W)\}\right] \right\}.\label{eq:gfps}
\]
where
\[
  u_a(W) \equiv \int_2^W p_a(Q|Q)dQ
               = a\log\left( \frac{\log W}{\log 2} \right).
\]

The final partons distribution is then obtained by inserting
Eq.~\eref{eq:gfps} into Eq.~\eref{eq:ffps} according to Eq.~\eref{eq:F}.

The clan parameters can then be calculated to be
\[
  \Nbar(W) = 1 + A \log\left( \frac{\log W}{\log 2} \right)
    \label{eq:Nfps}
\]
and
\[
  \nc(W) = \frac{1}{\Nbar(W)} \left\{1 - \frac{A}{a} \left[
            1 - \left( \frac{\log W}{\log 2} \right)^a \right]
	\right\}.
\]
As expected, the average number of clans depends only on the parameter $A$
which regulates the first step of the emission process; the average number
of partons in an {\em average} clan depends on both parameters (whereas
the distribution in a single clan of definite virtuality depends of
course only on $a$).
It should be noticed that in the limit of high initial virtuality
(energy), $W \to \infty$, the dependence of $\nc(W)$ on $A$ disappears.

\begin{figure}
\begin{center}\mbox{%
\epsfig{file=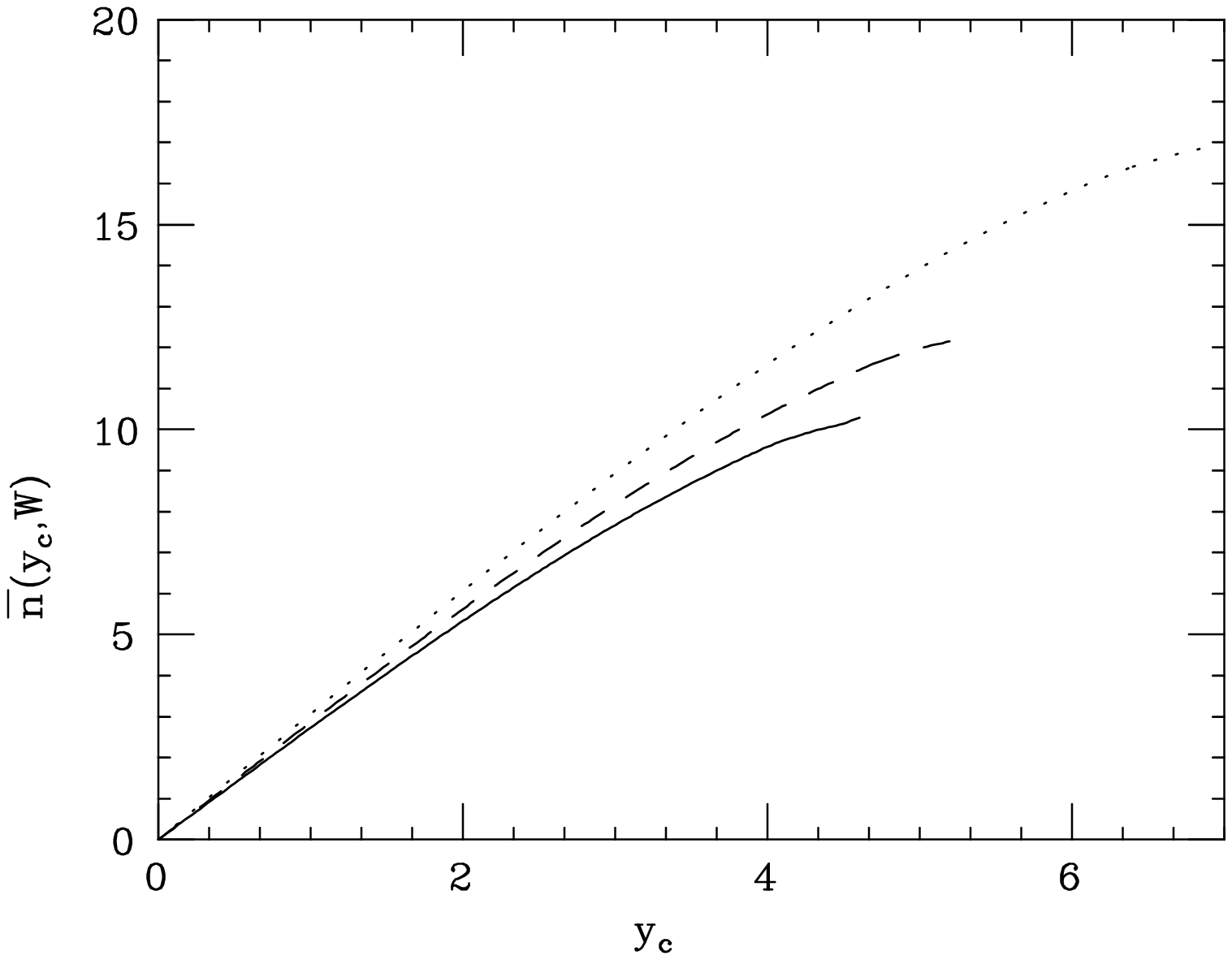,height=5.7cm,%
bbllx=86pt,bblly=227pt,bburx=525pt,bbury=571pt}\hspace{0.3cm}
\epsfig{file=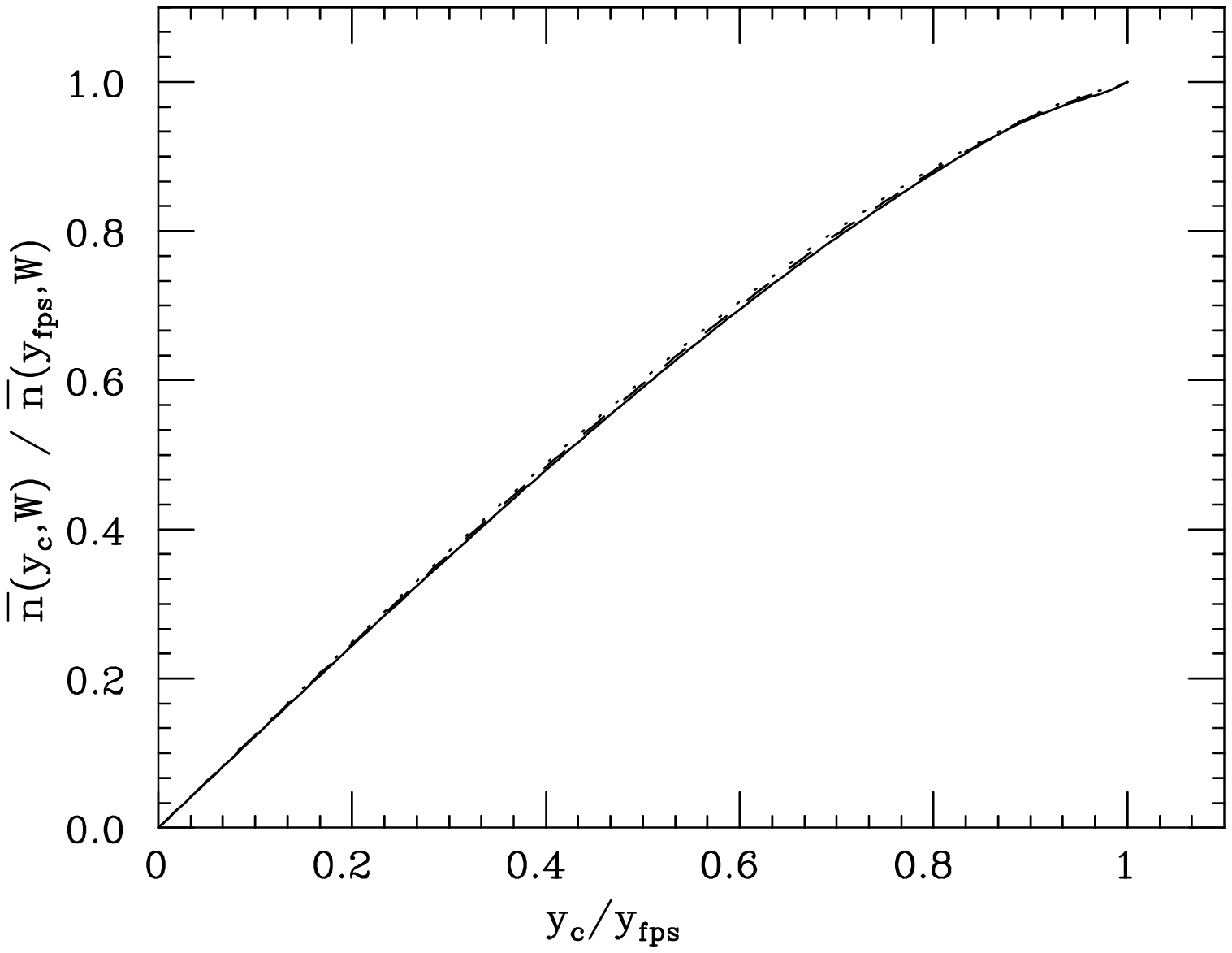,height=5.7cm,%
bbllx=86pt,bblly=227pt,bburx=525pt,bbury=571pt}
}\end{center}
\fcaption{{\em Left:} Average number of partons in the shower,
$\bar n(y_c,W)$,  as a function of the width of the
rapidity interval $y_c$ obtained analytically
in the GSPS model with $A$ = 2, $a$ = 1 at different maximum allowed
virtualities $W$ = 50 GeV (solid line), 100 GeV (dashed line) and 500 GeV
(dotted line).
{\em Right:} The same quantity is plotted normalized to the value
in full phase space as a function of the rescaled rapidity
interval.}\label{fig:avgn}
\end{figure}

The solution in rapidity intervals is, as explained in the previous
section, a convolution of Eq.~\eref{eq:ffps} with the MD of an average
clan; however, since we are here interested only on clan parameters properties,
we will calculate only these, and not the full distribution.

\begin{figure}\begin{center}
\mbox{\epsfig{file=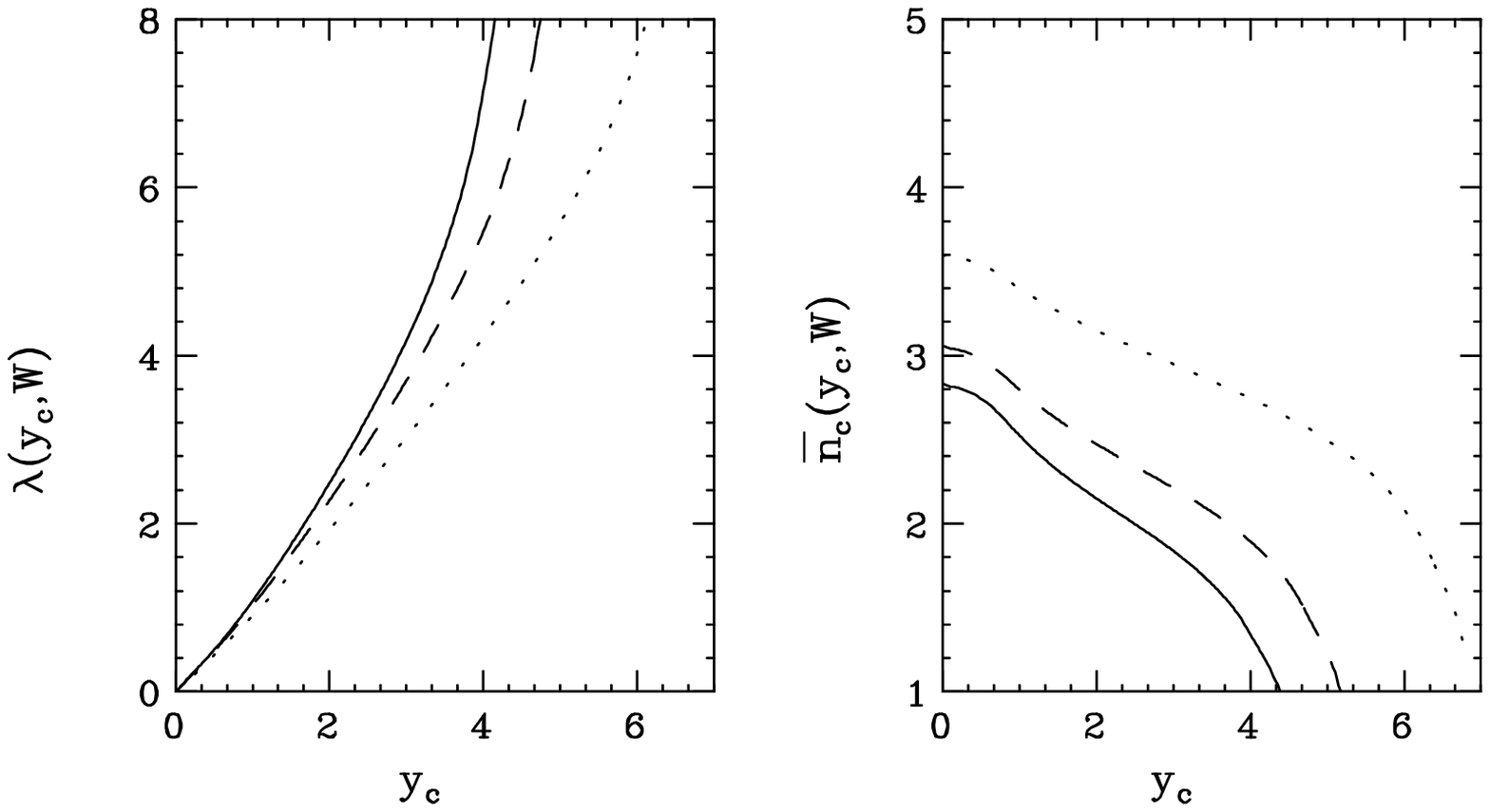,height=7cm,%
bbllx=44pt,bblly=193pt,bburx=504pt,bbury=444pt}}
\end{center}
\fcaption{Results of the GSPS model for the parameters
of clan analysis as a function
of the width of the
rapidity interval $y_c$ at
different maximum allowed virtualities $W$ = 50 GeV (solid line),
100 GeV (dashed line) and 500 GeV (dotted line).
Analytical solution with $A = 2$ and $a = 1$.}
\label{fig:lambda}
\end{figure}

The actual calculations in rapidity intervals
have been carried out for $A=2$ and $a=1$, values
which avoid  nasty
inessential  calculations and  make possible  the analytical
solution of the model without hurting its logic.

In Fig.~\ref{fig:avgn} we plot the average number of final partons
in both the standard form (on the left) and in a rescaled form
(to the right), showing an interesting scaling behaviour with energy.
This scaling in $W$ is found to depend on the parameter $a$, as different
values of $a$ give different scaling curves,
differently from the scaling found in\cite{GSPS}
for $\bar N(\Dy,W)/\bar N({\mathrm{fps}},W)$ which by its own
definition is independent of the mechanism at work inside clans.

In Fig.~\ref{fig:lambda} we plot the parameter $\lambda(\Dy,W)$, defined
by requiring that full distribution be a compound Poisson distribution,
and the corresponding average number of partons per clan. Remember that
is the parameter $\lambda$ which should be compared, {\em in rapidity
intervals not too close to full phase space}, with the average number
of clans as obtained from NBD fits.
One can clearly see in the figure the linear rise of $\lambda(\Dy,W)$
as $\Dy$ grows, and the slow decrease with energy at fixed $\Dy$.
The result for $\nc(\Dy,W)$ is reasonable but not fully satisfactory:
in the limit $\Dy\to 0$, $\nc(\Dy,W)$ points to constant values, which
differ at different energies, in contrast with the expected $W$-independent
value $\nc(0,W)=1$. The explanation for this anomaly lies in the approximation
we had to use in the analytical calculations, which fail in the very
small intervals $\Dy < 1$. Indeed a Monte Carlo version of the model
shows\cite{GSPS:2} that the slope of $\lambda(\Dy,W)$ when $\Dy\to 0$ matches
that of $\nbar(\Dy,W)$ in Fig.~\ref{fig:avgn}, so that actually one finds
$\nc(0,W)=1$.

%--------------------------------------------
\section{Conclusions}

We have illustrated the GSPS model and its results; it is a parton
shower model which was built from QCD-inspired splitting functions
in virtuality and in rapidity, with Sudakov form factor normalization,
to which the phenomenologically established idea of clans was added
by allowing at each step in the cascading local violations of the
energy-momentum conservation law (which is recovered globally in
a statistical sense). The model has important predictive power
in regions not presently accessible to full perturbative QCD;
the results on clan analysis are consistent with what we know
of single gluon jets disentangled using a jet finding algorithm
from the \jetset\ Monte Carlo program and analyzed at parton level
by assuming generalized local parton-hadron duality\cite{Single}.
These predictions can (and hopefully will) be tested in the near future
as clean samples of single gluon and quark jets of different
energies have been separated at LEP\cite{DEL:Brussel};
some preliminary data are available
in full phase space only, an analysis of
which is under way\cite{future}.
\ifpreprint\pagebreak[4]\fi

%--------------------------------------------
\section{Acknowledgements}

The authors would like to thank the organizers for the fruitful
atmosphere provided for this meeting.

%--------------------------------------------
\section{References}
%% References file: staralesna.ref
%% Creator: REFLATEX.AWK 1.0
%% Usage: \input this file where references should go
%%


\begin{thebibliography}{10}
%
\bibitem{AGLVH:1}  % 1
A. Giovannini and L. Van Hove, {\it Z.\ Phys.}{\bf\ C30} (1986) 391.

\bibitem{Buschbeck}  % 2
\ifpreprint B. Buschbeck, talk given at the XXV
International Symposium on Multiparticle Dynamics (Star\'a Lesn\'a, Slovakia,
12-16 Sept. 1995), to be published in the proceedings.
\else B. Buschbeck, these proceedings.\fi

\bibitem{Single}  % 3
F. Bianchi, A. Giovannini, S. Lupia and R. Ugoccioni,
{\it Z. Phys.}{\bf\ C58} (1993) 71.

\bibitem{DEL:4}  % 4
P.\ Abreu et al., DELPHI Collaboration, {\it Z.\ Phys.}{\bf\ C56} (1992) 63.

\bibitem{AGLVH:2}  % 5
L. Van Hove and A. Giovannini, {\it Acta Phys.\ Pol.}{\bf\ B19} (1988) 917.

\bibitem{GSPS}  % 6
R. Ugoccioni, A. Giovannini and S. Lupia,
   {\it Z.~Phys.}{\bf\ C64} (1994) 453.

\bibitem{GSPS:2}  % 7
A.~Giovannini, S.~Lupia and R.~Ugoccioni,
  ``The average number of partons per clan in rapidity intervals
  in parton showers'', DFTT 95-32, MPI-PhT/95-43, LU TP 95-11.
  To be published in {\it Z.~Phys.~C}.

\bibitem{AGLVH:4}  % 8
A. Giovannini and L. Van Hove, {\it Acta Phys.\ Pol.}{\bf\ B19} (1988) 495.

\bibitem{DEL:Brussel}  % 9
\ifpreprint G. Guy, talk given at the XXV
International Symposium on Multiparticle Dynamics (Star\'a Lesn\'a, Slovakia,
12-16 Sept. 1995), to be published in the proceedings.
\else G. Guy, these proceedings.\fi

\bibitem{future}  % 10
R. Ugoccioni, A. Giovannini, S. Lupia, in preparation.

\end{thebibliography}
\end{document}